# Indexing Properties of Primitive Pythagorean Triples for Cryptography Applications

Yashwanth Kothapalli
Oklahoma State University, Stillwater, OK-74078
yashwanth.kothapalli@okstate.edu

**Abstract**: This paper presents new properties of Primitive Pythagorean Triples (PPT) that have relevance in applications where events of different probability need to be generated and in cryptography.

## Introduction

The generation of events of specific probability is of importance in cryptography and in applications such as e-commerce [1],[2]. Such events may be generated by a variety of methods that include prime reciprocals [3],[4],[5], or by the use of specific modular operations [6]. The generation of random events is also tied up with the question of algorithmic probability [7]. Here we consider Pythagorean triples to generate probability events.

The lengths of sides of a right angled triangle represent a Pythagorean triple. A Pythagorean triple (a, b, c) should satisfy the condition $a^2 + b^2 = c^2$ as said in the pythogorous theorem which acts as the basis for trigonometry. We can generate infinity number of Pythagorean triples i.e. x (a, b, c) where x>1. A primitive Pythagorean triple is the one in which neither of the three numbers have any common factor. We can write a, b, c as

$$a = st$$

$$b = \frac{s^2 - t^2}{2}$$

$$c = \frac{s^2 + t^2}{2}$$

Where s, t are odd and co-primes to each other and $c + b = s^2$ and also $c - b = t^2$. For example for the triple (3, 4, 5) s=3 and t=1.

The Pythagorean triples a, b, c are divisible by either 3 or 4 or 5 separately or jointly so as the primitive Pythagorean triples. This property is used and divided all the PPTs into six classes by Kak [8], who also presents their historical background. If we could index sequences of these classes in different ways, this will have cryptographic applications. In this paper, the number of occurrences of transitions amongst pairs of classes in the first 10,000 PPTs is presented.

## Background

Euclidean Pythagorean primitive triples [9], [10] may be obtained using the formula

$$a = m^2 - n^2$$

$$c = m^2 + n^2$$

$$b = 2mn$$

Where m, n are relatively prime to each other and only of them is even and the other is odd.

We need to have pair of positive integers in which one of them is odd and the other is even and also relatively prime to each other in order to generate a PPT. We can do that by taking an odd number and writing it as sum of two numbers. If this pair of numbers happens to be relatively prime to each other, we can use them to generate a primitive Pythagorean triple. Once we get the PPT our next task is to find to which class they belong. Depending upon the divisibility of PPT (a, b, c) by 3 and 5 they have been classified into six different classes [8]:

1. Class A in which a is divisible by 3 and c divisible by 5

| div | a | b | c |
|---|---|---|---|
| 3 | X | | |
| 5 | | | X |

2. Class B in which a is divisible by 5 and b is divisible 3

| div | a | b | c |
|---|---|---|---|
| 3 | | X | |
| 5 | X | | |

3. Class C in which a is divisible by 3 and 5.

| div | a | b | c |
|---|---|---|---|
| 3 | | | X |
| 5 | | | X |

4. Class D in which b is divisible by 3 and c by 5

| div | a | b | c |
|---|---|---|---|
| 3 |   | X |   |
| 5 |   |   | X |

5. Class E in which a is divisible by 3 and b by 5

| div | a | b | c |
|---|---|---|---|
| 3 | X |   |   |
| 5 |   | X |   |

6. Class F in which is b divisible by 3 and 5

| div | a | b | c |
|---|---|---|---|
| 3 |   | X |   |
| 5 |   | X |   |

An explanatory example is given below in Table 1.

Table 1. Given an odd number, obtained from sum of co prime pairs, one can generate a PPT and find its class.

| Odd number | Co-prime pair | PPT generated | class |
|---|---|---|---|
| 3 | (2,1) | (3,4,5) | A |
| 5 | (3,2) | (5,12,13) | B |
|   | (4,1) | (15,8,17) | C |
| 7 | (4,3) | (7,24,25) | D |
|   | (5,2) | (21,20,29) | E |
|   | (6,1) | (35,12,37) | B |
| 9 | (5,4) | (9,40,41) | E |

|    | (7,2)  | (45,28,53)   | C |
|----|--------|--------------|---|
|    | (8,1)  | (63,16,65)   | A |
| 11 | (6,5)  | (11,60,61)   | F |
|    | (7,4)  | (33,56,65)   | A |
|    | (8,3)  | (55,48,73)   | B |
|    | (9,2)  | (77,36,85)   | D |
|    | (10,1) | (99,20,101)  | E |

Below is the list of the first 34 PPTs.

```
PPT            w(x)
(3,4,5)          A
(5, 12, 13)      B
(15, 8, 17)      C
(7, 24, 25)      D
(21, 20, 29)     E
(35, 12, 37)     B
(9, 40, 41)      E
(45, 28, 53)     C
(11, 60, 61)     F
(33, 56, 65)     A
(63, 16, 65)     A
(55, 48, 73)     B
(13, 84, 85)     D
(77, 36, 85)     D
(39, 80, 89)     E
(65, 72, 97)     B
(99, 20,101)     E
(91, 60,109)     F
(15,112,113)     C
(117, 44, 125)   A
(105, 88,137)    C
(17,144,145)     D
(143, 24,145)    D
(51,140,149)     E
(85,132,157)     B
(119,120,169)    F
(165, 52,173)    C
(19,180,181)     F
(57,176,185)     A
(153,104,185)    A
(95,168,193)     B
```

|     (195, 28,197)   C |
|     (133,156,205)   D |
|     (187, 84, 205)  D |

These are arranged in increasing order of hypotenuse length (c). If the hypotenuse lengths are equal then they placed according to the increasing order of 'a'. The corresponding sequence of classes for the above 34 PPTs is ABCDEBECFAABDDEBEFCACDDEBFCFAABCDD. This sequence does not include all possible pair wise transitions. To obtain all pair wise transitions we need to go to 8182 PPTs. If we take w(x)'s continuously 2, 3, 4 or 5 at a time there will be repetitions after first 26 PPTs. If we take i=6 then                                    will be 28 unique combinations. The graph of this behavior is given below.

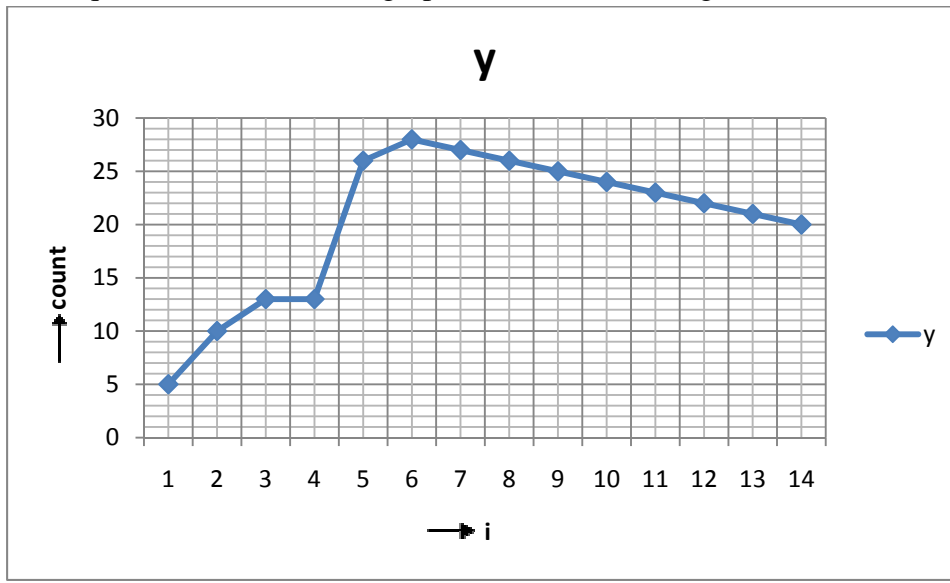

**Figure 1.** Graph between count of unique sequences and different values of i.

Here the x-axis is i in                                    , where as Y-axis gives the no: of unique combinations for that i.

**Property 1.**

For any prime number *y*, *n* number of co prime pairs are possible whose sum is the prime number itself through which we can generate *n* number of PPTs where            .

Proof: Since every number has *n* pairs of numbers whose is number itself. So, we need to show that for any prime number all the *n* pairs are co primes

Let *y* be a prime number and let            . Let's say that r, s are not co-primes and none of them are zero. Therefore there exists a number, let's say *z*, which divides both the numbers(r, s) and *z< y* since r, s are less than *y*. Since z divides both the r, s, it must also divide *y* which is a contradiction, since *y* is a prime number. Therefore we can say that all the pairs (r, s) for a prime number are co primes .There will be $2n + 1$ such pairs. Therefore we can generate 'n' number of PPTs out of a prime number *y* where $y = 2n + 1$.

**Property 2**

For all the odd multiples of 15, the PPTs generated will belong to class C.

**Property 3**

All the prime numbers which end with 3 or 7, except for 3, will follow a cyclic sequence DEBFADFCFDAFBED.

**Property 4**

All the prime numbers which end with 1 or 5, except for 5, will follow a cyclic sequence FABDEFDCDFEDBAF.

By observing two consecutive PPTs and drawing a table, we get the results of Table 2.

**Table 2.** Properties of first 10,000 PPTs arranged in increasing order of hypotenuse.

|   | A | | B | | C | | D | | E | | F | |
|---|---|---|---|---|---|---|---|---|---|---|---|---|
| A | X | y | X | y | X | y | X | y | X | y | X | y |
|   | 997 | 10 | 382 | 1 | 172 | 20 | 79 | 144 | 28 | 480 | 3 | 818 |
| B | X | y | X | y | X | y | X | y | X | y | X | y |
|   | 9 | 3016 | 471 | 79 | 406 | 2 | 177 | 12 | 403 | 6 | 196 | 25 |
| C | X | y | X | y | X | y | X | y | X | y | X | y |
|   | 180 | 19 | 120 | 171 | 487 | 60 | 371 | 3 | 171 | 61 | 336 | 8 |
| D | X | y | X | y | X | y | X | y | X | y | X | y |
|   | 2 | 5441 | 173 | 53 | 12 | 290 | 1006 | 13 | 446 | 4 | 36 | 769 |
| E | X | y | X | y | X | y | X | y | X | y | X | y |
|   | 32 | 389 | 353 | 5 | 199 | 7 | 2 | 7035 | 620 | 35 | 468 | 17 |
| F | X | y | X | y | X | y | X | y | X | y | X | y |
|   | 440 | 9 | 163 | 96 | 389 | 18 | 40 | 288 | 7 | 3338 | 623 | 65 |

Table 2 summarizes several properties for first 10,000 PPTs. Here X gives you the number of occurrences and y gives you the first occurrence. The sequence of PPTs will change when arranged by different conditions.

The table shows how the transition probabilities are different and will change with different indexing schemes. This will allow for different probability events to be generated.

## Indexing of PPTs

PPTs may be indexed in a variety of ways. The most obvious ones are listed below, where we also list the total length of the sequence in which all transitions between the six classes occur.

1. Arranged in increasing order of a in (a, b, c). All transitions occur in sequence length of 300. The sequence begins as: ABDEFDCCDFEABECEBAABDBFBECAFDDACDC.
2. Arranged in increasing order of b in (a, b, c). All transitions occur in sequence length of 4037. The sequence begins as: ACBBAEEDDCCDEABCAFFBEDDCACFBEDDBAF.

3. Arranged in increasing order of c-b in (a, b, c). All transitions occur in sequence length of 132. The sequence begins as: ABDEFDCDFCEAEEABCBBBBADFCFDEAADCDC.

4. Arranged in increasing order of c-a in (a, b, c). All transitions occur in sequence length of 504. The sequence begins as: ACBAEDCBECDACDBFDEABCAFEFDBDCEBDAF.

5. Arranged in increasing order of b-a in (a, b, c). All transitions occur in sequence length of 147. The sequence begins as: CDCDEAAADFBCCCBEAFBBDADEEBFDBECADF.

Other indexing schemes may be used by placing conditions on the three parameters or by imposing uniform or nonuniform decimation of the specific PPT sequence.

If only the sender and the recipient are in the know regarding the indexing scheme used, then the corresponding PPT term would be the key that the two parties would have successfully exchanged.

## Conclusion

This paper presents some new results on the sequence of PPTs (a, b, c) related to their six classes *w(x)*. It shows how different sequences are obtained using different indexing schemes. The PPT sequence which is arranged in increasing order of $c - b$ covers all possibilities, when pairs of classes are considered, in fewer steps compared to other basic PPT sequences.